\documentclass[aps,prx,twocolumn,superscriptaddress,showpacs,longbibliography]{revtex4-1}
\usepackage{amsmath}
\usepackage{graphicx}
\usepackage{amsfonts}
\usepackage{verbatim}
\usepackage{amssymb}
\usepackage{color}
\usepackage{dsfont}
\usepackage{amsmath} 
\usepackage{hyperref}
\hypersetup{colorlinks = true, urlcolor = blue, linkcolor = blue, citecolor = blue}

\newcommand{\su}{\uparrow} 
\newcommand{\sd}{\downarrow} 
\newcommand{\bpm}{\begin{pmatrix}}
\newcommand{\epm}{\end{pmatrix}}

\newcommand{\nn}{\nonumber \\}

\newcommand{\rN}{\rm{N}}
\newcommand{\rF}{\rm{F}}
\newcommand{\rS}{\rm{S}}

\begin{document}


\title{Optimizing the topological properties of stacking semiconductor-ferromagnet-superconductor heterostructures}
\author{Chun-Xiao Liu}\email{Electronic address: chunxiaoliu62@gmail.com}
\affiliation{Qutech and Kavli Institute of Nanoscience, Delft University of Technology, Delft 2600 GA, The Netherlands}

\author{Michael Wimmer}
\affiliation{Qutech and Kavli Institute of Nanoscience, Delft University of Technology, Delft 2600 GA, The Netherlands}

\date{\today}

\begin{abstract}
We study the electronic properties of a planar semiconductor-superconductor heterostructure, in which a thin ferromagnetic insulator layer lies in between and acts as a spin filtering barrier. 
We find that in such a system one can simultaneously enhance the strengths of all the three important induced physical quantities, i.e., Rashba spin-orbit coupling, exchange coupling, and superconducting pairing potential, for the hybrid mode by external gating.
Our results show specific advantage of this stacked device geometry compared to conventional devices.
We further discuss how to optimize geometrical parameters for the heterostructure and complement our numerical simulations with analytic calculations.
\end{abstract}

\maketitle


\section{Introduction}
Topological superconductivity can emerge in a spin-orbit coupled semiconductor-superconductor heterostructure when the induced Zeeman spin splitting is larger than the induced superconducting gap~\cite{Nayak2008Non-Abelian, Alicea2012New, Leijnse2012Introduction, Beenakker2013Search, Stanescu2013Majorana, Jiang2013Non, Elliott2015Colloquium, DasSarma2015Majorana, Sato2016Majorana, Sato2017Topological, Aguado2017Majorana, Lutchyn2018Majorana, Zhang2019Next, Frolov2020Topological,Sau2010Generic, Lutchyn2010Majorana, Oreg2010Helical, Sau2010NonAbelian}.
In a recent experiment, following the theoretical proposals~\cite{Sau2010Generic, Sau2010NonAbelian}, experimentalists added a thin ferromagnetic film on the heterostructure to induce a Zeeman spin splitting via exchange coupling, and indeed tunnel spectroscopy signatures of topological superconductivity have been observed in the absence of an external magnetic field~\cite{Vaitiekenas2020Zerobias}.
More interestingly, the experimental hints at topological superconductivity only appeared in devices of a particular geometry, i.e., when the ferromagnetic insulator and superconductor layers partially overlap, but not in the non-overlapping counterparts.
This indicates the crucial role of device geometry in determining the electronic properties of the heterostructure.
As pointed out in Refs.~\cite{Liu2021Electronic, Escribano2021Tunable, Khindanov2021Topological, Woods2020Electrostatic}, the relative positioning of the ferromagnetic and superconducting layers on top of the nanowire could affect the electrostatic potential profile inside the semiconductor as well as the total strength of the induced Zeeman spin splitting in a delicate manner.
Additionally, it has also been shown via a phenomenological model that topological superconductivity can appear in a hybrid system when the ferromagnetic film is inserted as a spin-filtering barrier separating the semiconductor and superconductor layers~\cite{Maiani2021Topological, Langbehn2021Topological}.
The advantage of such a proposal using ferromagnetic spin-filtering layer is the generic applicability to both quasi-one-dimensional nanowire and two-dimensional-electron-gas systems. 
Here, we consider a microscopic model of the device geometry as shown in Fig.~\ref{fig:schematic}, going beyond the phenomenological model and aiming to address the following two key questions: First, whether it is possible to realize topological superconductivity in the device with realistic physical parameters. Second, how to optimize the topological properties of the heterostructure using experimental knobs, e.g., by gating or varying the geometrical parameters.
The answers to the above two questions will be valuable to the future exploration of topological materials and devices.
In order to obtain a faithful description of the electronic properties in the heterostructure, we adopt the method of microscopic device simulation, in which all the three material layers are treated on equal footing and the electrostatic potential profile inside the semiconductor is calculated by the self-consistent Thomas-Fermi-Poisson equation.

\section{Model Hamiltonian and method}

\begin{figure}
\begin{center}
\includegraphics[width=\linewidth]{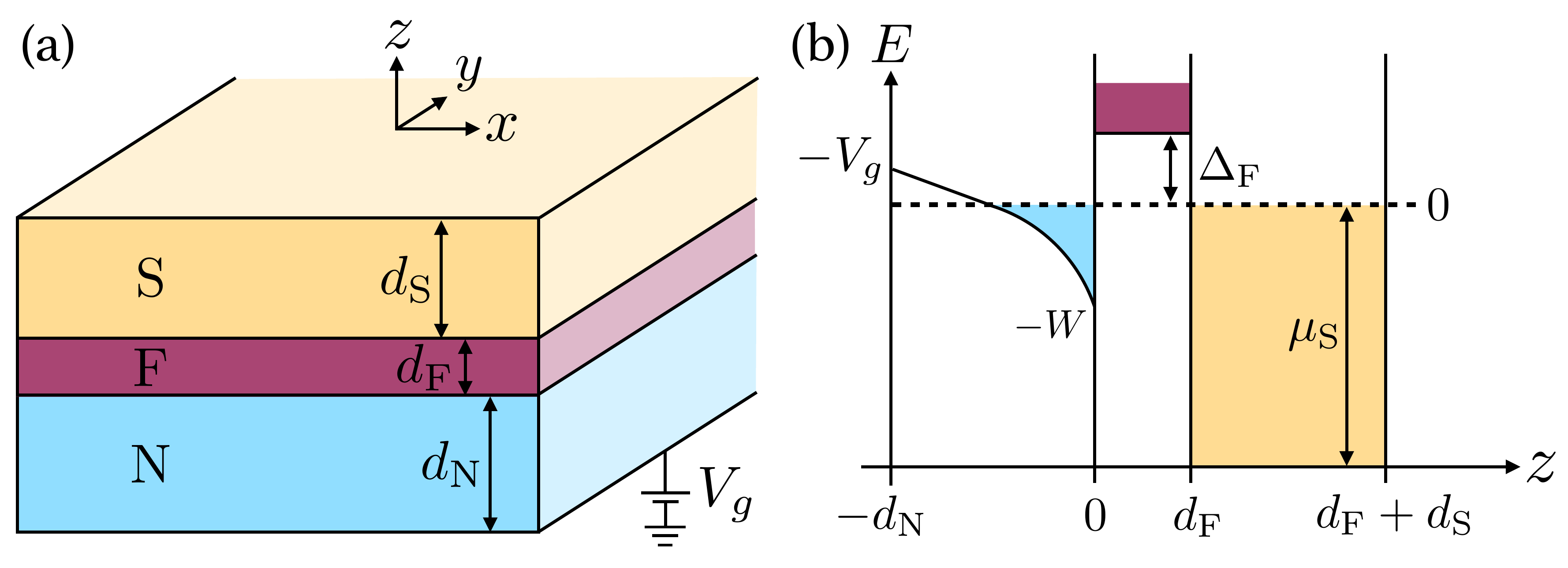}
\caption{(a) Schematic of the semiconductor-ferromagnet-superconductor heterostructure studied in the current work.
The material layers stack along the $z$-axis and extend in the $xy$-plane.
The electrostatic environment inside the semiconductor is determined jointly by the semiconductor-ferromagnet band offset $W$ and the bottom gate voltage $V_g$.
(b) Bandedge profile of the conduction-band electrons in the heterostructure.}
\label{fig:schematic} 
\end{center}
\end{figure}

The hybrid system we consider in this work is a planar semiconductor-ferromagnet-superconductor (N-F-S) heterostructure.
The material layers stack along the $z$-axis and extend infinitely in the $xy$-plane, as indicated by the schematic in Fig.~\ref{fig:schematic}(a).
The semiconductor, ferromagnetic insulator and superconductor layers occupy the region of $-d_{\rm{N}} < z < 0$, $0 < z < d_{\rm{F}}$, and $d_{\rm{F}} < z < d_{\rm{F}} + d_{\rm{S}}$ respectively.
In our model, we assume translational symmetry along the $y$-axis, and quasi-translational symmetry along the $x$-axis.
That is, we consider a superlattice along the $x$-axis, and impose a periodic boundary condition on the superlattice boundaries.
The purpose of such a superlattice modeling is to include the effect of disorder in the 2D cross section. 
Although such a 2D modeling of disorder is only an approximation of the realistic 3D situation, it already captures the important effect of disorder on the semiconductor-superconductor heterostructure~\cite{Winkler2019Unified}. 
The Bogoliubov-de Gennes (BdG) Hamiltonian for the heterostructure is
\begin{align}
H_{\rm{BdG}} &= \frac{\hbar^2}{2} \left( \frac{ -\partial^2_x + k^2_y}{m^*(z)}  -\partial_{z} \frac{1}{m^*(z)} \partial_z + E_b(z) \right) \tau_z \nn
& - \alpha_{R}(z) \left( i \partial_x \sigma_y + k_y \sigma_x \right)\tau_z + h(z) \sigma_z \nn
& + \Delta(z) \tau_x + U(x,z) \tau_z,
\label{eq:H_BdG}
\end{align}
in the basis of $(\psi_{e \su}, \psi_{e \sd}, \psi_{h \sd}, -\psi_{h \su})$. 
$\hbar$ is the Planck constant, $\sigma_{x,y,z}$, $\tau_{x,y,z}$ are Pauli matrices acting on the spin and Nambu spaces, and $\hbar k_y$ is the momentum along the $y$-axis.
$m^*(z)$ is the effective mass of the conduction-band electrons, with 
\begin{align}
m^*(z) =
\begin{cases}
m_{\rN}=0.023 m_{\rm{e}}, & z \in \rN \\
m_{\rF}=0.3 m_{\rm{e}} & z \in \rF \\
m_{\rS}= m_{\rm{e}} & z \in \rS.
\end{cases}       
\end{align}
Here the values for the effective masses are chosen according to InAs, EuS and Al, with $m_{\rm{e}}$ being the rest electron mass~\cite{Winkler2003Spin, Xavier1967On, Cochran1958Superconducting}.
$E_b(z)$ is the conduction band edge profile, with 
\begin{align}
E_b(z) =
\begin{cases}
-e \phi(z) & z \in \rN \\
\Delta_{\rF} & z \in \rF \\
-\mu_{\rS} & z \in \rS.
\end{cases}       
\end{align}
Here $\Delta_{\rF} = 0.5$ eV is the insulating gap above the chemical potential in the ferromagnetic insulator~\cite{Liu2020Coherent, Cochran1958Superconducting}, and $\mu_{\rS}=11$ eV is the Fermi energy of the S layer.
$\phi(z)$ is the electrostatic potential in N, and is obtained by solving the one-dimensional self-consistent Thomas-Fermi-Poisson equation in the semiconductor layer
\begin{align}
\partial^2_z \phi(z)  = \frac{\rho[ \phi(z) ]}{ \varepsilon },
\label{eq:TFP}
\end{align}
with $\varepsilon=15.5\varepsilon_0$ being the dielectric constant of InAs.
The boundary conditions for the Poisson equation are $\phi(0)=W \approx 0.3$ V~\cite{Liu2021Electronic} and $\phi(-d_{\rN})=V_g$, which are determined jointly by the semiconductor-ferromagnet band offset $W$ at the top and the gate voltage $V_g$ at the bottom of the N layer.
In reality, there is another thin dielectric layer between the semiconductor and the backgate, leading to a marginal difference in the electrostatic potential simulation inside the semiconductor. 
Thus our simplified modeling of the effect of the backgate is justified~\cite{Mikkelsen2018Hybridization}.
Here the charge density $\rho$ includes both the conduction-band electrons and the valence-band holes inside the semiconductor, i.e.,
\begin{align}
&\rho[\phi(z)]= \rho_{\text{e}}(\phi) + \rho_{\text{hh}}(\phi) + \rho_{\text{lh}}(\phi), \nn
&\rho_{\text{e}}(\phi) =  -\frac{e}{3\pi^2 } \left( \frac{2m_{\rN} e \phi \theta(\phi)}{\hbar^2}  \right)^{3/2}, \nn
&\rho_{\text{hh}/\text{lh}}(\phi) =  \frac{e}{3\pi^2} \left( \frac{ 2m_{\text{hh}/\text{lh}} (-e\phi-E_0) \theta(-e\phi-E_0)}{\hbar^2}  \right)^{3/2},
\end{align}
where $e>0$ is the elementary charge unit, $m_{\text{hh}}=0.41~m_{\rm{e}}$, $m_{\text{lh}}=0.026~m_{\rm{e}}$ are the heavy- and light-hole effective mass in unit of electron mass, $E_0=0.418~$eV is the band gap between conduction and valence bands, and $\theta(x)$ is the Heaviside step function. 
$\alpha_R(z)$ is the strength of Rashba spin-orbit coupling which is finite only in N. Its strength is determined by the local electric field
\begin{align}
\alpha_R(z) =  
\frac{eP^2}{3} \left( \frac{1}{E^2_0} - \frac{1}{(E_0+\Delta_0)^2} \right) \partial_z \phi(z),\quad  z \in \rN
\label{eq:SOC}
\end{align}
from the eight-band $k \cdot p$ theory, with $P=0.9197$ eV nm, and $\Delta_0=0.38$ eV~\cite{Winkler2003Spin}.
$h(z)$ is the strength of the exchange field in F with 
\begin{align}
h(z) =  h, \quad z \in \rF,
\end{align}
and zero elsewhere. 
We assume that the exchange field points along the $z$-axis and the field strength is a variable to be studied later, although the estimated strength in EuS is 100 meV $< h < $ 200 meV. 
$\Delta(z)$ is the pairing potential in the conventional $s$-wave superconductor layer~\cite{Cochran1958Superconducting}
\begin{align}
\Delta(z) = \Delta=0.35~\mathrm{meV}, \quad z \in \rS.
\end{align}
Additionally, in a realistic semiconductor-superconductor hybrid material, the S layer is inevitably disordered, e.g., owing to the oxidization of the outer S layer or the formation of grain domains in the process of material growth.
Here we model such a disorder effect by including a 2D disorder potential $U(x,z)$ in S, which is effectively introducing Fermi energy fluctuations. 
Such a fluctuation is assumed to be spatially uncorrelated, i.e.,
\begin{align}
\langle U(x,z) U(x',z') \rangle = U^2_D \delta(x-x') \delta(z-z'), \quad z, z' \in \rS,
\end{align}
with $\delta(x)$ being the delta function. $U_D$ is the fluctuation amplitude, and $\langle \cdot \rangle$ denotes disorder averaging.

To implement the numerical calculation, we first discretize the continuum Hamiltonian in Eq.~\eqref{eq:H_BdG} on a square lattice with lattice constant $a=0.1$ nm using Kwant~\cite{kwant}, and then calculate the eigenenergies and eigenwavefunctions by diagonalizing the sparse-matrix Hamiltonian.

\begin{figure}[t]
\begin{center}
\includegraphics[width=\linewidth]{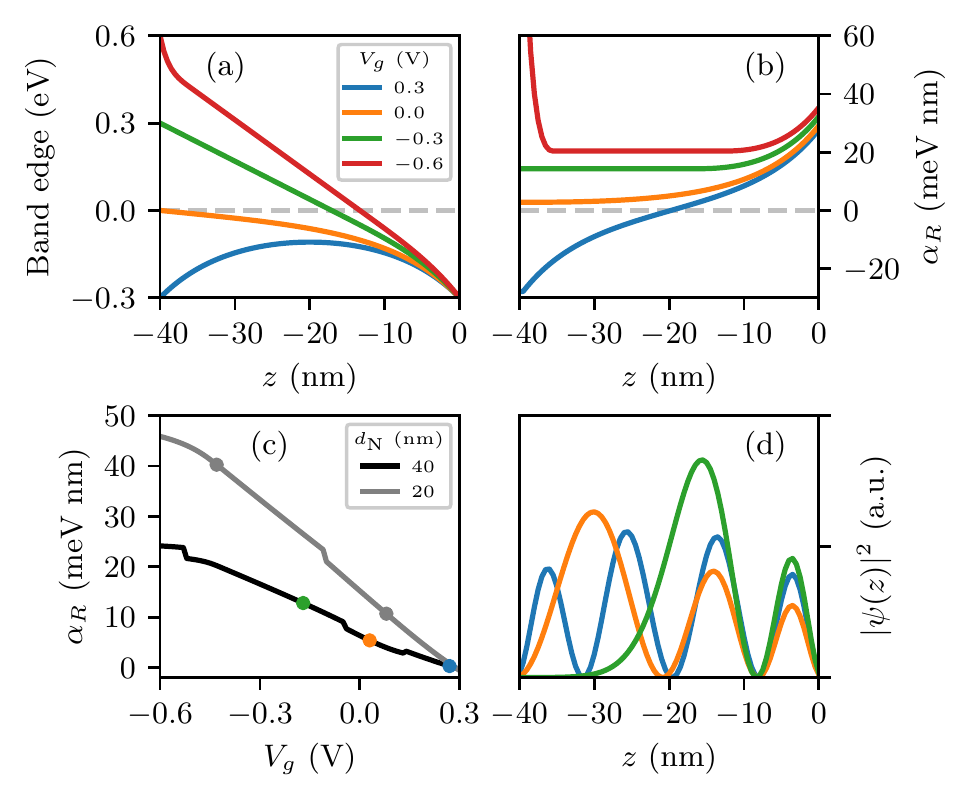}
\caption{(a) and (b) Profiles of band edge and spin-orbit coupling strength in a 40 nm-thick bare N layer for different gate voltages.
(c) Strength of spin-orbit coupling for the particular eigenstate closest to Fermi surface ($E=0$) as a function of gate voltage.
Jumps in the curve indicate the switch between adjacent eigenstates, and dots denote the eigenstates with $E=0$.
(d) Wavefunction profiles of the eigenstates corresponding to the colored dots in (c) for the 40 nm-thick bare N layer.
}
\label{fig:bare_N} 
\end{center}
\end{figure}

\section{Bare semiconductors}
We first consider a bare semiconductor layer, focusing on the electrostatic potential profile and the corresponding spin-orbit coupling strength profile in real space.
We include the role of the F layer only as the interface band offset $W$ at the top boundary of the bare N layer in the self-consistent Thomas-Fermi-Poisson calculation.
Then with the potential and spin-orbit coupling strength profiles serving as input, we solve for the eigenstates in a bare N layer and study the dependence of the state-specific charge density and the spin-orbit coupling strength on the layer thickness ($d_{\rN}$) and gate voltages ($V_g$).

In Figs.~\ref{fig:bare_N}(a) and~\ref{fig:bare_N}(b), we show the spatial profiles of the electrostatic potential $\phi(z)$ and the corresponding spin-orbit coupling strength $\alpha_R(z)$ in a bare N layer of thickness $d_{\rN}=40$ nm for different values of backgate voltages.
For a positive backgate voltage, charge accumulation wells show up at both the top (N-F interface at $z=0$) and the bottom (N-substrate interface at $z=-40$ nm) of the layer, but with an opposite sign of spin-orbit coupling strengths.
Otherwise, a more negative gate voltage would increase the slope of the electrostatic potential profile, thereby enhancing the spin-orbit coupling strength.
In addition, due to the screening effect of accumulation electrons in the semiconductor, the strength of spin-orbit coupling is stronger at the N-F interface than in the bulk.

With the calculated electrostatic potential and spin-orbit coupling strength profile as input, we then obtain the eigenenergies and eigenfunctions in the bare N layer by solving the following normal Hamiltonian 
\begin{align}
&H_{\rm{N}} \psi^{(n)}(x, z) =E^{(n)} \psi^{(n)}(x, z), \nn
&H_{\rm{N}} = -\frac{\hbar^2  }{2 m_{\rN}}\left( \partial^2_x + \partial^2_z \right)  - e \phi(z) -i \alpha_{R}(z) \partial_x \sigma_y,
\label{eq:H_N}
\end{align}
where $E^{(n)}$ and $\psi^{(n)}(x, z) $ are the eigenenergies and eigenfunctions for $H_{\rm{N}}$. 
The state-specific spin-orbit coupling strength is defined as
\begin{align}
\alpha^{(n)}_{R} = \sum_{\sigma = \su, \sd} \int dx dz \alpha_{R}(z) | \psi^{(n)}_{\sigma}(x, z) |^2.
\end{align}
The curves in Fig.~\ref{fig:bare_N}(c) show the strength of the spin-orbit coupling for the eigenstate closest to the Fermi surface as a function of gate voltage.
It shows that a more negative gate voltage tends to enhance the spin-orbit coupling strength for the particular eigenstate.
This is because a more negative backgate voltage would increase the electric field across the material, and at the same time pushes the electronic wavefunction towards the N-F interface [see Fig.~\ref{fig:bare_N}(d)], where the local spin-orbit coupling strength is the strongest.
Additionally, the gray-color curve in Fig.~\ref{fig:bare_N}(c) shows that a thinner N layer would induce an even stronger spin-orbit coupling strength for the eigenstates at the cost of allowing fewer transverse modes within the same gate voltage range.

\section{N-F-S heterostructures}
We now consider a semiconductor-ferromagnet-superconductor heterostructure.
The peculiarity of the hybrid system is that all the three material layers are stacking on top of the adjacent layer and that a thin ferromagnetic insulator layer is inserted between the semiconductor and superconductor layers, as shown in the schematic in Fig.~\ref{fig:schematic}(a).
In this section, we aim to figure out the effects of gate voltages and geometrical parameters on the electronic and topological properties in the hybrid system.

\begin{figure*}[t]
\centering
\includegraphics[width= 0.75\textwidth]{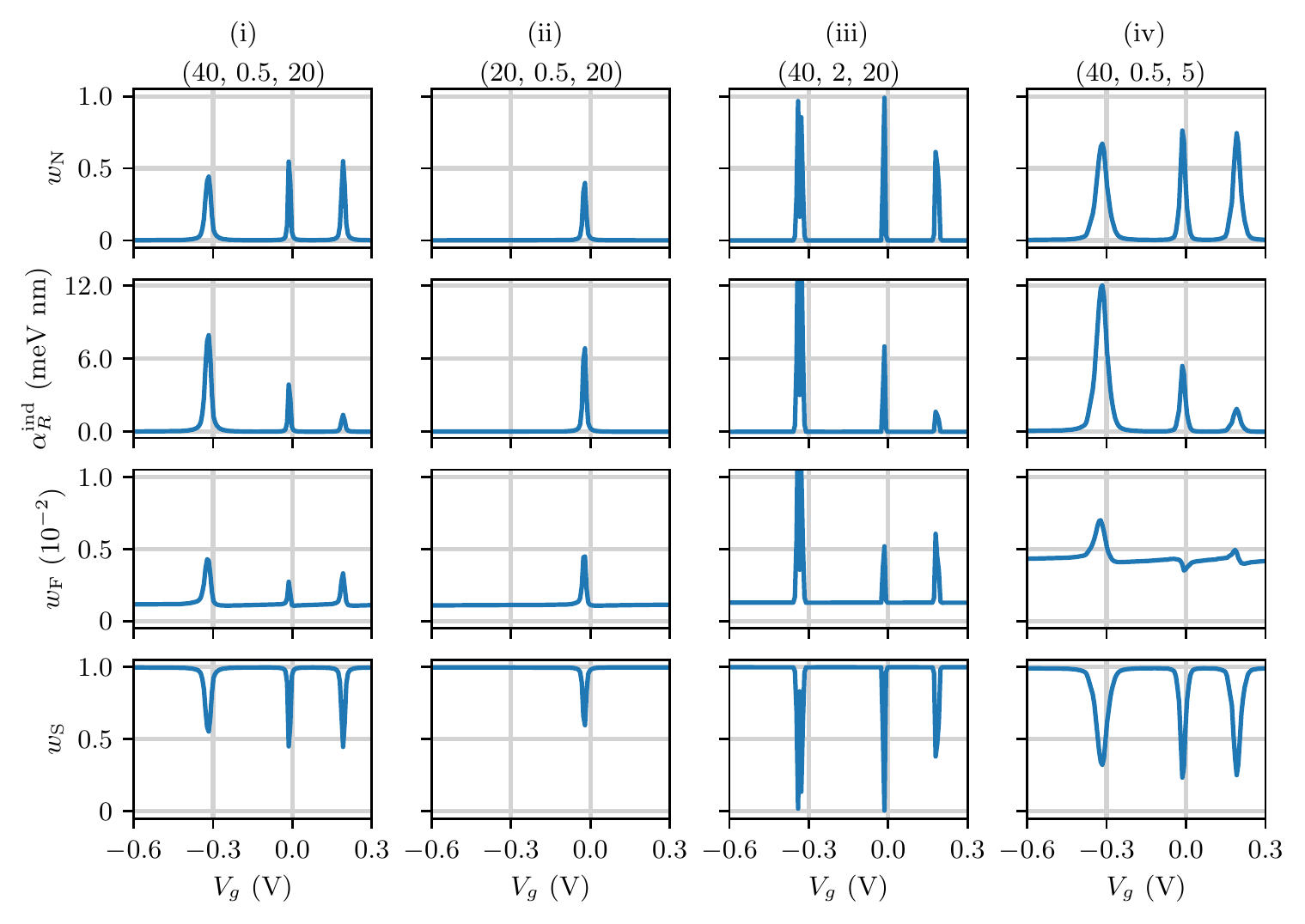}
\caption{Electronic properties, i.e., disorder averaged $w_{\rN/\rF/\rS}$ and $\alpha^{\rm{ind}}_{R}$, of N-F-S heterostructures with different geometrical parameters.
Each column has the same parameter set of $(d_{\rN},d_{\rF},d_{\rS} )$ in unit of nm.
For other physical parameters, we choose $h=100$ meV, $k_y=0$, and $U_D=0.5$ eV.
}
\label{fig:w_NFS}
\end{figure*}

\subsection{Electronic properties}
For the hybrid states which are potentially responsible for topological superconductivity, they obtain the induced spin-orbit coupling, exchange coupling, and superconducting pairing potential by distributing their wavefunctions inside the semiconductor, ferromagnetic insulator, and superconductor layers, respectively.
We thus define the weight of an eigenstate inside the particular material layer as follows
\begin{align}
w_{i} = \sum_{\sigma = \su, \sd} \sum_{\eta=\rm{e}, \rm{h}} \int_{(x, z) \in \Omega_i} dx dz |\psi_{\sigma \eta}(x,z)|^2,
\end{align}
where $i=$N, F, or S, and $\Omega_{i}$ denotes the space occupied by the corresponding material layer.
Normalization of the wavefunction constrains that 
\begin{align}
w_{\rN} + w_{\rF} + w_{\rS} = 1. 
\end{align}
In the parameter regime the current work focuses on, i.e., $h \ll \Delta_{\rF}$ and $\Delta \ll \mu_{\rS}$, the induced exchange coupling and superconducting pairing potential have a direct proportionality with the weights in F and S layers, i.e.,
\begin{align}
h^{\rm{ind}} \approx w_{\rF} h, \quad \Delta^{\rm{ind}} \approx w_{\rS} \Delta,
\label{eq:h_Delta_ind}
\end{align}
with $h$ and $\Delta$ being the bare values.
On the other hand, for the induced spin-orbit coupling strength, we need to take into account its spatial dependence and define 
\begin{align}
\alpha^{\rm{ind}}_R \approx \sum_{\sigma = \su, \sd} \sum_{\eta=\rm{e}, \rm{h}} \int_{(x, z) \in \Omega_{\rN}} dx dz \alpha_R(z) |\psi_{\sigma \eta}(x,z)|^2.
\end{align}
In Fig.~\ref{fig:w_NFS}, we show the calculated four quantities $w_{\rN/\rF/\rS}$ and $\alpha^{\rm{ind}}_R$ of the eigenstate closest to the Fermi energy ($E=0$) as a function of gate voltage ($V_g$) for different combinations of geometrical parameters ($d_{\rN},d_{\rF}, d_{\rS} $).
In order to eliminate the dependence on a particular disorder configuration and to make our findings more generic, all these quantities are averaged over 150 different disorder realizations in the model calculation.  
Furthermore, we choose the amplitude of the disorder potential fluctuation to be $U_D = 0.5$ eV, so that the hybridization between the semiconductor and superconductor is strong enough and that the simulation results do not depend on the precise value of $U_D$.

Column-(i) of Fig.~\ref{fig:w_NFS} shows the calculated $w_{i}$ and $\alpha^{\rm{ind}}_R$ for an N-F-S heterostructure with geometrical parameters $(d_{\rN},d_{\rF},d_{\rS}) = (50, 0.5, 20)$ in unit of nm. 
Here the thickness of F layer is chosen to be comparable to the penetration length in the insulator, i.e., $d_{\rF} \sim \lambda_{\rF}= \hbar/\sqrt{2m_{\rF} \Delta_{\rF}}$.
The resonance peaks and dips in the curves in Fig.~\ref{fig:w_NFS} denote the presence of a semiconductor-superconductor hybrid state in the heterostructure, with its peak and dip values indicating the corresponding strength of the physical quantities.
In contrast, those relatively flat curves represent states essentially localized in the superconducting layer, because they have $w_{\rS} \sim 1$ and $w_{\rN} \sim 0$.
Here both $w_{\rF}$ and $w_{\rS}$ of the hybrid mode increase when the applied backgate voltage becomes more negative, because the wavefunctions are pushed towards the F and S side by the applied electric field.
Surprisingly, although $w_{\rN}$ decreases in this gating process, the strength of the induced spin-orbit coupling is still increasing, since a stronger spin-orbit coupling strength in N (see Fig.~\ref{fig:bare_N}) compensates the loss of its weight.
Thereby we see the simultaneous enhancement of all the three important quantities $\alpha^{\rm{ind}}_R$, $h^{\rm{ind}}$, $\Delta^{\rm{ind}}$ in the N-F-S heterostructure by external gating.

From column-(ii) to column-(iv) in Fig.~\ref{fig:w_NFS}, we show the calculated $w_{i}$ and $\alpha^{\rm{ind}}_R$ for heterostructures with different geometrical parameters, in order to see how the thickness of each material layer affects the wavefunction distributions and the induced quantities.
In column-(ii), we set the N layer thickness $d_{\rN} = 20$ nm as compared to $d_{\rN} = 40$ nm in column-(i).
Although the optimized induced quantities ($\alpha^{\rm{ind}}_{R} \sim 7 $ meV nm, $w_{\rF} \sim 0.5 \times 10^{-2}$, $w_{\rS} \sim 0.5$) are similar to those in column-(i), the number of available hybrid modes within the voltage range is reduced from three to only one owing to a narrower confinement in the N layer.
On the other hand, when the F layer thickness is larger than its decay length ($d_{\rF} \gg \lambda_{\rF}$), as shown in column-(iii) of Fig.~\ref{fig:w_NFS}, the semiconductor and superconductor layers become decoupled.
Especially for the modes at $V_g \sim -0.3$ V and $V_g \sim 0.0$ V, they have $w_{\rN} \sim 1$ and $w_{\rS} \sim 0$, which means that the semiconductor states cannot be proximitized with superconductivity when the ferromagnetic insulator barrier is too thick. 
Finally, if we reduce the thickness of S layer, we will find that both the strength of the spin-orbit coupling and the exchange coupling of the hybrid modes increase but at the cost of a reduced proximity superconducting gap. 
Note that the hybrid-state peaks and dips in column-(iv) are much broader than those in the other columns.
This is because the level spacing in a thinner S layer becomes larger ($\delta E \propto d^{-2}_{\rS}$), thus giving rise to a bigger statistical fluctuations in the disorder averaging. 
Furthermore, as shown in the $w_{\rF}$ plot of column-(iv), the induced exchange coupling for the superconducting states (flat curves beside the resonance peaks or dips) increases. 
An enhanced exchange coupling in the superconducting states will reduce the BCS gap, thus giving a weaker topological protection.

\begin{figure*}[t]
\centering
\includegraphics[width= 0.75\textwidth]{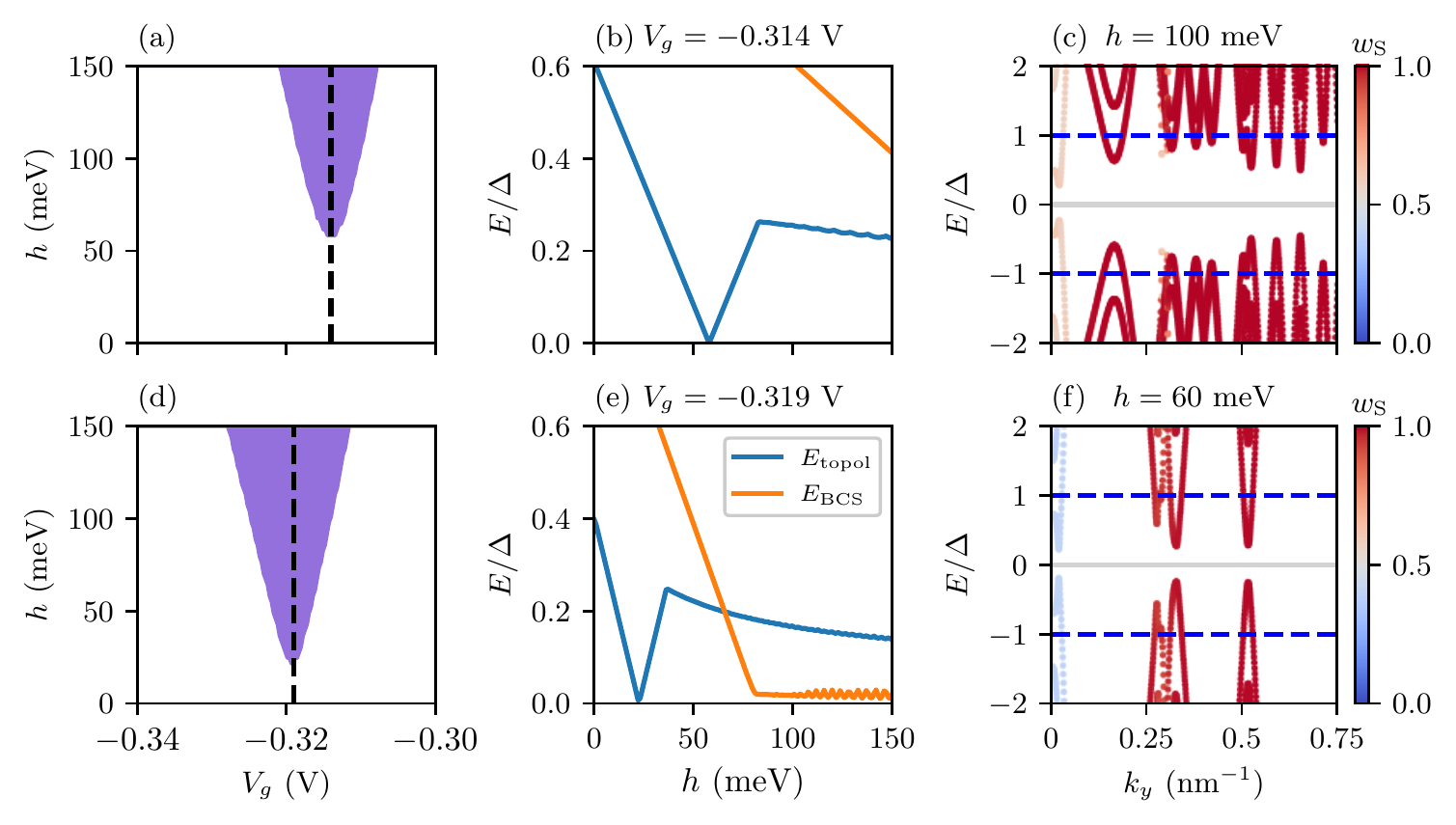}
\caption{topological phase diagram, gaps, and band diagrams of N-F-S heterostructures.
The geometrical parameters are $(d_{\rN}, d_{\rF}, d_{\rS} ) = (40, 0.5, 20)$ in unit of nm for the upper panel, and (40, 0.5, 5) for the lower panel.
}
\label{fig:topol_phase}
\end{figure*}

\subsection{Topological phase diagrams and gaps}
We now consider the topological phase diagrams and the topological gaps for an N-F-S heterostructure.
The key questions to address in this subsection include whether topological superconductivity can be realized with realistic physical parameters and how to optimize the topological properties via external gating and geometrical parameters.
Based on the numerical results in the previous subsection, we first choose the hybrid state at $V_g \sim -0.3$ V for $(d_{\rN}, d_{\rF}, d_{\rS}) =$ (40 nm, 0.5 nm, 20 nm), because it has a strong semiconductor-superconductor hybridization as well as large induced spin-orbit coupling strength and exchange coupling.
The corresponding topological phase diagram in the ($V_g$, $h$) plane is shown in Fig.~\ref{fig:topol_phase}(a).
The purple area denotes the topological phase, while the white area the trivial phase.
For the optimal value of gate voltage, i.e., $V^*_g \sim -0.314$ V, the topological phase transition takes place at about $h \sim 60$ meV, which is smaller than the typical exchange coupling in a ferromagnetic insulator material, e.g., $h_{\rm{EuS}} \sim 150$ meV~\cite{Mauger1986The}.
In Fig.~\ref{fig:topol_phase}(b), we further give the evolution of the excitation gap as a function of $h$ at the optimal gate voltage.
We find that the topological gap of the hybrid state [blue line in Fig.~\ref{fig:topol_phase}(b)] can be larger than 0.2 $\Delta$ in large $h$.
In addition, as indicated by the orange line in Fig.~\ref{fig:topol_phase}(b), as $h$ in the F layer increases, it will also reduce the BCS gap in the superconductor.
Here the BCS gap ($E_{\rm{BCS}}$) is defined as the lowest excitation energy for states essentially localized inside the superconducting layer ($w_{S}\approx 1$).
However, in this scenario, since the BCS gap is always larger than the topological gap, it does not affect the topological properties of the heterostructure in an essential way.
For comparison, we also show the topological phase diagram and gap evolution for the heterostructure with a thinner S layer ($d_{\rN} = 5$ nm), because as shown in Fig.~\ref{fig:w_NFS}, this would enhance the induced spin-orbit coupling and exchange coupling at the cost of a smaller induced superconducting gap.
Figures~\ref{fig:topol_phase}(d) and~\ref{fig:topol_phase}(e) show that the minimal critical exchange coupling $h^* \sim 25$ meV indeed decreases compared to the thick S layer scenario.
However, the topological gap is now smaller than 0.2 $\Delta$ at large $h$, and what's more the BCS gap closes at $h \sim 80$ meV, thus making the S layer transition to normal metal.
Thus for a thinner S layer, although the topological phase transition requires a weaker exchange coupling strength, both the topological and BCS gaps in the hybrid system are reduced.

\section{Analytic understanding}
\begin{figure}[t]
\centering
\includegraphics[width= 0.5\textwidth]{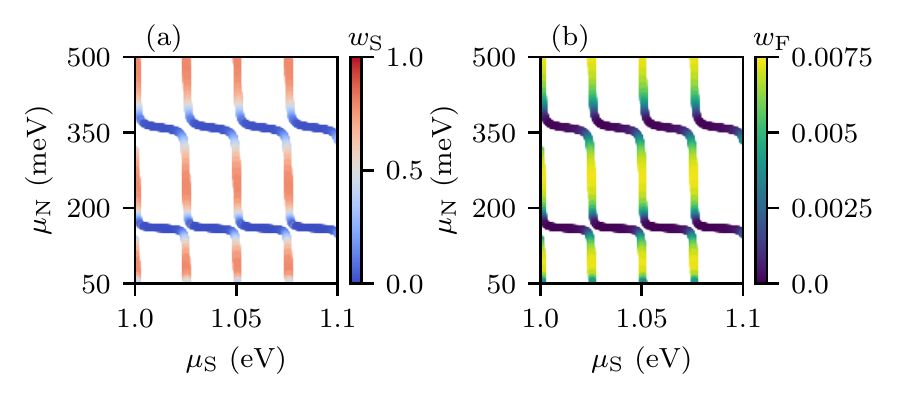}
\caption{Traces of $E=0$ for a one-dimensional spinless model of N-F-S heterostructure with geometrical parameters $(d_{\rN}, d_{\rF}, d_{\rS} ) = (20, 0.3, 50)$ in unit of nm.
}
\label{fig:analytic}
\end{figure}

In this section, we use the wavefunction approach to understand the electronic properties of an N-F-S heterostructure.
We consider a one-dimensional spinless double quantum well model for studying the wavefunction distribution in the hybrid system~\cite{Mikkelsen2018Hybridization, Reeg2018Metallization}.
In the parameter regime of our interest, the spin-orbit coupling in N, the exchange field in F, and the superconductivity in S only play a perturbative role in determining the wavefunction distribution.
Thereby we use the following simplified model Hamiltonian for our analytic calculation
\begin{align}
H = \frac{\hbar^2}{2}  \left( - \partial_z\frac{1}{m(z)} \partial_z + \frac{k^2_x + k^2_y}{m(z)} \right) + V(z).
\end{align}
Here $V(z)$ is the conduction bandedge profile, with 
\begin{align}
V(z) =
\begin{cases}
-\mu_{\rN}, & -d_{\rN} < z <0 \\
\Delta_{\rF}, & 0< z < d_{\rF} \\
-\mu_{\rS}, & d_{\rF} < z < d_{\rF} + d_{\rS},
\end{cases}
\end{align}
where we model the N and S layers as square quantum wells and the F layer as an insulating barrier between them.
We are particularly interested in solving for the eigenstates at the Fermi energy, i.e.,
\begin{align}
H(k_x=0, k_y=0)\psi(z) =0,
\end{align}
with $\psi(z)$ being the wavefunction along the stacking direction.
The ansatz of the bound-state wave function is
\begin{align}
\psi(z) = 
\begin{cases}
A \sin (k_{\rN}(z+d_{\rN})), & -d_{\rN} < z <0 \\
B e^{-\kappa_{\rF} z} + C e^{\kappa_{\rF}(z - d_{\rF})}, & 0 < z < d_{\rF} \\
-\sin ( k_{\rS} (z - d_{\rF} - d_{\rS} ) ), & d_{\rF} < z < d_{\rF} + d_{\rS},
\end{cases}  
\label{eq:wavefunction}
\end{align}
with $k^2_{\rN}  = 2 m_{\rN} \mu_{\rN} / \hbar^2$, $\kappa_{\rF}^2  = 2 m_{\rF} \Delta_{\rF} / \hbar^2$, $k^2_{\rS}  = 2 m_{\rS} \mu_{\rS} / \hbar^2$.
By matching the wave function and its derivative at $z=0$ and $z=d_{\rF}$, we have
\begin{align}
& A = \frac{ \kappa_{\rF} \sin \phi_{\rS} \cosh \phi_{\rF} + k_{\rS} \cos \phi_{\rS} \sinh \phi_{\rF} }{\kappa_{\rF} \sin \phi_{\rN}}, \nn
& B =  e^{\phi_{\rF}} \frac{ \kappa_{\rF} \sin \phi_{\rS} + k_{\rS} \cos \phi_{\rS} }{2 \kappa_{\rF}} , \nn
& C = \frac{ \kappa_{\rF} \sin \phi_{\rS} - k_{\rS} \cos \phi_{\rS} }{2 \kappa_{\rF}},
\end{align}
where the dimensionless phases are $\phi_{\rN} = k_{\rN} d_{\rN}$, $\phi_{\rF} = \kappa_{\rF} d_{\rF}$, $\phi_{\rS} = k_{\rS} d_{\rS}$.
The transcendental equation determining the bound state solution is
\begin{align} 
A k_{\rN} \cos \phi_{\rN} + B \kappa_{\rF} - C \kappa_{\rF} e^{-\phi_{\rF}} = 0.
\end{align}
Figure~\ref{fig:analytic} shows the traces of $E=0$ states in the ($\mu_{\rS}, \mu_{\rN}$) plane and the weights in different parts.
In Fig.~\ref{fig:analytic}(a), the red vertical lines denote the S-like states, while those dark blue horizontal lines denote the N-like states, and the anti-crossings indicate the N-S hybridization.
Both the energy levels and the weights are periodic functions of $\mu_{\rS}$, which means the calculated physical quantities in the clean limit depends sensitively on the Fermi energy in the superconductor.
Such a Fermi energy dependence, which is not observed in realistic devices, can be eliminated by performing a disorder averaging in calculating the physical quantities.
For the S-like states, the analytic form of the wavefunction can be obtained by setting $A \to 0, B \to 0$ in Eq.~\eqref{eq:wavefunction}, because the fraction of wavefunction in the N layer is negligible.
The wavefunction now is 
\begin{align}
\psi(z) = 
\begin{cases}
\sin (\phi_{\rS}) e^{\kappa_{\rF} (z - d_{\rF})}, & 0 < z < d_{\rF} \\
-\sin( k_{\rS}(z - d_{\rF} - d_{\rS} )), & d_{\rF} < z < d_{\rF} + d_{\rS},
\end{cases}  
\label{eq:s_like}
\end{align}
in the regime of $\kappa^{-1}_{\rF} = \lambda_{\rF} \lesssim d_{\rF}$. 
The corresponding weight in F is
\begin{align}
w_{\rF} &= \frac{\int^{d_{\rF}}_0 dz |\psi(z)|^2}{\int^{d_{\rF}}_0 dz |\psi(z)|^2 + \int^{d_{\rF} + d_{\rS}}_{d_{\rF}} dz |\psi(z)|^2} \nn
& \approx  \frac{\sin^2(\phi_S) \lambda_{\rF} }{ d_{\rS} - \sin (2 \phi_{\rS}) / 2k_{\rS} } \approx \frac{ \lambda_{\rF} }{ 2 d_{\rS} },
\end{align}
where in the last line we take the average value for the sinusoidal functions in a $\mu_{\rS}$-period, i.e., $\sin^2(\phi_S) \to 1/2$ and $ \sin(2\phi_S) \to 0$.
Thereby using $h^{\mathrm{ind}} \approx w_{\rF} h$ in Eq.~\eqref{eq:h_Delta_ind}, we can estimate the BCS gap by
\begin{align}
E_{\rm{BCS}} \approx \Delta - h^{\rm{ind}} \approx \Delta - h \lambda_{\rF} / 2 d_{\rS},
\end{align}
where we assume that for $s$-wave superconductivity, the BCS gap is reduced linearly by the induced exchange coupling.
By setting $E_{\rm{BCS}} =0$ we immediately have $h^* = 2 \Delta d_{\rS} / \lambda_{\rF} $ to be the strength of the critical exchange coupling for closing the BCS gap.
This analytic result is consistent with the numerical simulations shown in Figs.~\ref{fig:topol_phase}(b) and~\ref{fig:topol_phase}(e), where the BCS gap for a thinner S layer closes at a much smaller exchange coupling strength.

\section{Discussions}
Here we elaborate on how to choose the thickness of S and F layers for optimal electronic and topological properties in a semiconductor-superconductor heterostructure with a spin-filtering ferromagnetic layer.
In general, a thicker S layer would induce a larger topological gap as well as a larger BCS gap in the heterostructure, thus giving a clearer experimental signature and a better topological protection.
This is in contrast with semiconductor-superconductor heterostructures in the presence of an externally applied magnetic field.
There a thinner S layer is always preferred because the depairing effect from the orbital effect of magnetic field can be mitigated only in a superconductor with reduced cross section area.
On the other hand, since the effective Zeeman spin splitting in N-F-S heterostructures is induced by the exchange coupling inside the F layer, the F film is expected to have an optimal thickness such that a finite and strong exchange coupling can be induced in the hybrid material but without decoupling the semiconductor-superconductor hybridization.
We estimate the optimal F thickness to be around the corresponding penetration length, i.e., $d_{\rF} \sim \lambda_{\rF}$, with $\lambda_{\rF}$ being extracted from tunneling conductance measurements.
As for the N thickness, there is always a tradeoff between a larger number of available states within the backgate voltage range (thicker N layer) and stronger spin-orbit coupling strength (thinner N layer).
Since a systematic investigation on the relation between the semiconductor thickness/diameter and the strength of Rashba spin-orbit coupling is still missing, an experimental study even on a bare semiconductor layer would be very useful to understanding the electronic properties.
Finally, gating is an important knob in tuning the electronic properties of an N-F-S heterostructure. 
Here we find that a hybrid state obtained by a more negative gating generically has a better electronic properties for topological superconductivity, i.e., larger induced superconducting gap, stronger induced Zeeman splitting and a stronger strength of spin-orbit coupling.
Therefore, varying the backgate voltage to a more negative value would be a good choice for finding topological states with optimized properties and protection.

\section{Conclusions}
In conclusion, we have studied the electronic properties of a stacking N-F-S heterostructure with planar geometry.
By microscopic device simulation, we show that topological superconductivity can be realized in these devices with realistic physical parameters.
We also figure out how gating and varying the geometrical parameters could optimize the electronic and topological properties of the hybrid states in the device.
Although the focus of this work is on the planar device geometry, all the analysis and findings should carry over to nanowire-based devices as well, as long as a thin ferromagnetic layer lies in between the semiconductor and superconductor layers.

\acknowledgements
We are grateful to Anton Akhmerov and  Yu Liu for useful discussions.
This work is supported by a subsidy for top consortia for knowledge and innovation (TKl toeslag).

C.-X.L. conceived and designed the project, with the scope being later refined by M.W.
C.-X.L performed the numerical and analytic calculations.
All authors discussed the results and contributed to the writing of the manuscript.


\bibliography{references}






\end{document}